\documentclass[aps,twocolumn,floatfix,showpacs,amsmath,amsfonts]{revtex4-1}

\usepackage{graphicx}
\usepackage{bm}
\usepackage{units}
\usepackage{color}

\begin{document}

\title{$K$ dependent exchange interaction of the $1S$ ortho exciton in Cu$_2$O}

\author{Frank Schweiner}
\author{J\"org Main}
\author{G\"unter Wunner}
\affiliation{Institut f\"ur Theoretische Physik 1, Universit\"at Stuttgart,
  70550 Stuttgart, Germany}
\author{Christoph Uihlein}
\affiliation{Experimentelle Physik 2, Technische Universit\"at Dortmund, 44221 Dortmund, Germany}
\date{\today}

\begin{abstract}
When treating the exchange interaction of Wannier excitons,
usually only the leading terms of the
analytic and the nonanalytic exchange interaction
are considered. However, higher order terms can lead 
to a splitting of exciton states,
for which reason a splitting of the $1S$~exciton in cuprous oxide 
$\left(\mathrm{Cu_{2}O}\right)$ depending on its total momentum $\hbar K$
has been attributed to a $K$ dependent analytic exchange interaction
by Dasbach~\emph{et~al} [Phys.~Rev.~Lett.~\textbf{91}, 107401 (2003)].
Going beyond the common treatment of the exchange interaction, 
we derive the correct expressions for these $K$ dependent higher order terms
using $\boldsymbol{k}\cdot\boldsymbol{p}$ perturbation theory.
We prove that the appearance of a $K$ dependent exchange interaction
is inseparably connected with a $K$ independent exchange interaction of $P$ and $D$ excitons.
We estimate the magnitude of these terms for $\mathrm{Cu_{2}O}$
from microscopic calculations and show that they are far too small
to explain the observed $K$ dependent splitting. Instead, this splitting
has to be treated in terms of the dispersion of the excitons.
Furthermore, we prove the occurence of a coupling 
between longitudinal and transverse excitons in $\mathrm{Cu_{2}O}$
due to the $K$ dependent nonanalytic exchange interaction.

\end{abstract}


\maketitle

\section{Introduction~\label{sec:Introduction}}

Excitons are the quanta of the fundamental optical excitations in
both insulators and semiconductors in the visible and ultraviolet
spectrum of light. They consist of a negatively charged electron
in the conduction band and a positively charged hole in the valence
band. Wannier excitons extend over a huge
number of unit cells and can be described within
the simple band model as hydrogen-like particles~\cite{TOE}. Recently, the
corresponding hydrogen-like exciton absorption spectrum could be followed
up to a principal quantum number of $n=25$ in cuprous oxide $\left(\mathrm{Cu_{2}O}\right)$~\cite{GRE}.
This recent experiment led to a variety of new theoretical and experimental
investigations on the topic of excitons in $\mathrm{Cu_{2}O}$~\cite{QC,74,75,76,50,28,80,100,77}.

When investigating exciton spectra of $\mathrm{Cu_{2}O}$ using high resolution spectroscopy
and crystals of high quality, two of the most striking experimental findings 
are the observation of $F$ excitons and a splitting 
of the $1S$~exciton depending on its total momentum $\hbar K$.
Both effects cannot be understood within a simple effective mass model. 
Therefore, the $K$ dependent splitting of the $1S$ exciton was attributed by 
Dasbach~\emph{et~al}~\cite{9_1,8,9} to a $K$ dependent exchange interaction.
This is beyond the scope of the common treatment
of the exchange interaction for $\mathrm{Cu_{2}O}$, where
only a $K$ independent analytic exchange and a vanishing 
nonanalytic exchange interaction are considered.

Since we have recently shown that the observed splitting
could also be explained by taking full account of the 
anisotropic dispersion of the $\Gamma_5^+$ orbital
Bloch states~\cite{100}, 
we believe that the influence of a $K$
dependent exchange interaction on the $1S$ ortho exciton in 
$\mathrm{Cu_{2}O}$ deserves a closer investigation
as it is \emph{a priori} unknown whether dispersion \emph{and} 
exchange interaction are of the same size.

Although a preliminary investigation as regards the 
presence of a $K$ dependent exchange interaction 
was undertaken by Kavoulakis~\emph{et~al}~\cite{1},
their treatment was limited 
to the nonanalytic part of the exchange interaction only
and lacking a consideration of the complete
valence band structure of $\mathrm{Cu_{2}O}$.
Using $\boldsymbol{k}\cdot\boldsymbol{p}$ perturbation
theory, we derive general expressions for both the
analytic and nonanalytic part of the
exchange interaction for all direct excitons up to
basically arbitrary order in $K$.
This allows us not only to show the unknown fact
that the appearance of a $K$ dependent exchange interaction
is inseparably connected to a $K$ independent 
exchange interaction of $P$ and $D$ excitons
but also to estimate the magnitude 
of the $K$ dependent terms
from microscopic calculations for both parts of the interaction.
This is furthermore in contrast to
the simple group theoretical treatment
of the exchange interaction of Refs.~\cite{9_1,8,9}, which leads 
to $K$ dependent terms of the correct form 
but does not yield the their prefactors.
Since every $K$ dependent energy
as regards states of the symmetry $\Gamma_{5}^{+}$ must lead to matrices
of the form presented in Ref.~\cite{8}, the unambiguous assigment
of the experimentally observed $K$ 
dependent splitting to the exchange interaction is not
possible by these means.

Moreover, as regards the nonanalytic exchange interaction, we go beyond
the treatment of Kavoulakis~\emph{et~al}~\cite{1} and pay special 
attention to its angular dependency. 
This allows us to prove the
occurence of a coupling between longitudinal and transverse excitons 
in $\mathrm{Cu_{2}O}$ due to the $K$ 
dependent terms of this part of the exchange interaction.
Hence, we show that all three ortho exciton states couple to light
if the wave vector is not oriented in a direction of high symmetry.

The paper is organized as follows: In Sec.~\ref{sec:Exchange-interaction}
we discuss the exchange interaction of Wannier excitons
and derive the expressions for the $K$ dependent terms
of the analytic and the nonanalytic exchange energy.
Having pointed out the specific properties of excitons in 
$\mathrm{Cu_{2}O}$ in Sec.~\ref{sec:Excitons-in},
we investigate in Sec.~\ref{sub:Analytic-exchange-interaction}
the analytic and in Sec.~\ref{sub:Nonanalytic-exchange-interaction}
the nonanalytic exchange interaction
for the $1S$ exciton of cuprous oxide as well as the coupling between
longitudinal and transverse excitons.
Finally, we give a short summary and outlook in Sec.~\ref{sec:Summary-and-outlook}.

\section{Exchange interaction~\label{sec:Exchange-interaction}}

In this section we derive the $K$ dependent terms 
of the analytic and the nonanalytic exchange interaction
based on the main expressions
of the exchange interaction given in Refs.~\cite{E2,E3,TOE,1,SO}.
Within the scope of the simple band model the wave function
of an exciton is given by
\begin{eqnarray}
\Psi_{vc\,\nu\boldsymbol{K}} & = & \sum_{\boldsymbol{q}}f_{vc\,\nu}\left(\boldsymbol{q}\right)\Phi_{vc}^{\sigma\tau}\left(\boldsymbol{q}-\gamma\boldsymbol{K},\,\boldsymbol{q}+\alpha\boldsymbol{K}\right).\label{eq:Psiexc}
\end{eqnarray}
The envelope function $f_{vc\,\nu}\left(\boldsymbol{q}\right)$ is the Fourier transform of the 
hydrogen-like solution $F_{vc\,\nu}\left(\boldsymbol{\beta}\right)$ of the Wannier equation~\cite{TOE_5,TOE},
\begin{equation}
f_{vc\,\nu}\left(\boldsymbol{q}\right)=\frac{1}{\sqrt{N}}\sum_{\boldsymbol{\beta}}F_{vc\,\nu}\left(\boldsymbol{\beta}\right)e^{-i\boldsymbol{q}\boldsymbol{\beta}},
\end{equation}
with $\nu$ being a short
notation for the three quantum numbers $n$, $L$, and $M$. 
Note that the coordinate
$\boldsymbol{\beta}$ is a lattice vector which takes in general only discrete
values.
The
constant factors $\alpha=m_{\mathrm{e}}/(m_{\mathrm{e}}+m_{\mathrm{h}})$ and 
$\gamma=1-\alpha$ depend on the effective masses
of electron and hole. Additionally, the wave function~(\ref{eq:Psiexc}) contains
a Slater determinant of Bloch functions with one electron being
in a Bloch state of the conduction band and $N-1$ electrons in Bloch
states of the valence bands,
\begin{eqnarray}
& & \Phi_{vc}^{\sigma\tau}\left(\boldsymbol{k}_{\mathrm{h}},\,\boldsymbol{k}_{\mathrm{e}}\right) = \nonumber\\
& & \mathcal{A}\psi_{v\boldsymbol{k}_{1}\alpha}\psi_{v\boldsymbol{k}_{1}\beta}\cdots\psi_{v\boldsymbol{k}_{\mathrm{h}}\sigma}\psi_{c\boldsymbol{k}_{\mathrm{e}}\tau}\cdots\psi_{v\boldsymbol{k}_{N}\beta}.\label{eq:slater}
\end{eqnarray}
Here $\mathcal{A}$ denotes the antisymmetrization operator.

In the Wannier equation the exchange energy is missing since it is
often treated as a correction to the hydrogen-like solution~\cite{TOE}.
In general, the exchange energy between two exciton states $\Psi_{vc\,\nu\boldsymbol{K}}$
and $\Psi_{v'c'\,\nu'\boldsymbol{K}'}$ reads~\cite{TOE,E2}
\begin{widetext}
\begin{align}
E_{\mathrm{exch}}\left(vc\,\nu\boldsymbol{K},\, v'c'\,\nu'\boldsymbol{K}'\right) = &\: \delta_{\sigma\tau}\delta_{\sigma'\tau'}\delta_{\boldsymbol{K},\boldsymbol{K}'}\sum_{\boldsymbol{q},\boldsymbol{q}'}f_{vc\,\nu}^{*}\left(\boldsymbol{q}\right)f_{v'c'\,\nu'}\left(\boldsymbol{q}'\right)\nonumber \\
\nonumber \\
 \times & \int\mathrm{d}\boldsymbol{r}_{1}\,\int\mathrm{d}\boldsymbol{r}_{2}\,\psi_{c\boldsymbol{q}}^{*}\left(\boldsymbol{r}_{1}\right)\psi_{v\boldsymbol{q}-\boldsymbol{K}}\left(\boldsymbol{r}_{1}\right)\frac{e^{2}}{4\pi\varepsilon_{0}\varepsilon\left|\boldsymbol{r}_{1}-\boldsymbol{r}_{2}\right|}\psi_{c'\boldsymbol{q}'}\left(\boldsymbol{r}_{2}\right)\psi_{v'\boldsymbol{q}'-\boldsymbol{K}'}^{*}\left(\boldsymbol{r}_{2}\right).\label{eq:Hexchk}
\end{align}
\end{widetext}
The exchange energy includes the term $\delta_{\sigma\tau}\delta_{\sigma'\tau'}$.
Introducing the total spin $S=S_{\mathrm{e}}+S_{\mathrm{h}}=\tau-\sigma$
of electron and hole, this term can be written
with singlet and triplet states as $2\delta_{S,0}$~\cite{TO}. 

Inserting the Fourier transform~\cite{E3,SST}
\begin{equation}
\frac{1}{r}=\frac{4\pi}{NV_{\mathrm{uc}}}\sum_{\boldsymbol{G}}\sum_{\boldsymbol{k}\in\mathrm{BZ}}\,\frac{1}{\left(\boldsymbol{k}+\boldsymbol{G}\right)^{2}}e^{i\left(\boldsymbol{k}+\boldsymbol{G}\right)\boldsymbol{r}}\label{eq:1rFourier}
\end{equation}
with the volume of one unit cell of the lattice $V_{\mathrm{uc}}$ and reciprocal lattice vectors $\boldsymbol{G}$
in Eq.~(\ref{eq:Hexchk}), we can write the exchange energy as
\begin{eqnarray}
E_{\mathrm{exch}} & = & 2\delta_{S,0}\delta_{\boldsymbol{K},\boldsymbol{K}'}\nonumber\\
 & & \nonumber \\
 & \times & \sum_{\boldsymbol{G}}\frac{m_{vc\,\nu}^{*}\left(\boldsymbol{K},\,\boldsymbol{G}\right)m_{v'c'\,\nu'}\left(\boldsymbol{K},\,\boldsymbol{G}\right)}{\varepsilon_{0}\varepsilon V_{\mathrm{uc}}\left(\boldsymbol{K}+\boldsymbol{G}\right)^{2}}\label{eq:Hexchm2}
\end{eqnarray}
with
\begin{eqnarray}
m_{vc\,\nu}\left(\boldsymbol{K},\,\boldsymbol{G}\right) & = & \frac{e}{\sqrt{N}}\sum_{\boldsymbol{q}}\, f_{vc\,\nu}\left(\boldsymbol{q}\right)\nonumber\\
 & & \nonumber \\
 & \times & \left\langle u_{v\boldsymbol{q}-\gamma\boldsymbol{K}}\left|e^{-i\boldsymbol{G}\boldsymbol{r}}\right|u_{c\boldsymbol{q}+\alpha\boldsymbol{K}}\right\rangle .
\end{eqnarray}
The functions $u_{n\boldsymbol{k}}\left(\boldsymbol{r}\right)$ denote the lattice-periodic
part of the Bloch functions $\psi_{n\boldsymbol{k}}\left(\boldsymbol{k}\right)=e^{i\boldsymbol{k}\boldsymbol{r}}u_{n\boldsymbol{k}}\left(\boldsymbol{r}\right)$~\cite{SST}.
In the representation of Eq.~(\ref{eq:Hexchm2})
the exchange energy can be divided into the nonanalytic part
$E_{\mathrm{exch}}^{\mathrm{NA}}$, which is the summand with $\boldsymbol{G}=\boldsymbol{0}$,
and the analytic part $E_{\mathrm{exch}}^{\mathrm{A}}$, which is
the sum of the remaining terms.
Note that if the exchange energy is formulated in the
Wannier representation~\cite{TOE} instead of the representation with Bloch
functions, it is generally separated into a long-range
and a short-range part.
However, according to Refs.~\cite{E3,E2,E3_22} there is no identity between the
nonanalytic exchange and the long-range part or between the
analytic exchange and the short-range part
but only a close correspondence.

In the limit $Ka\ll1$ one obtains the simple expression~\cite{E2,TOE}
\begin{eqnarray}
E_{\mathrm{exch}}^{\mathrm{NA}} & = & 2\delta_{S,0}\delta_{\boldsymbol{K},\boldsymbol{K}'}\frac{1}{\varepsilon_{0}\varepsilon V_{\mathrm{uc}} K^2}\left(\boldsymbol{\mu}_{vc\,\nu\boldsymbol{K}}^{*}\boldsymbol{K}\right)\left(\boldsymbol{\mu}_{v'c'\,\nu'\boldsymbol{K}}\boldsymbol{K}\right)\nonumber\\
\nonumber\\
& + & \mathcal{O}\left(K^{2}a^{2}\right)\label{eq:HexchnaLR}
\end{eqnarray}
for the nonanalytic exchange energy of excitons in a cubic crystal. By $a$ we denote the lattice constant of the solid.
The expression~(\ref{eq:HexchnaLR}) depends only on the two
angles between $\boldsymbol{K}$ and the dipole moments $\boldsymbol{\mu}_{vc\,\nu\boldsymbol{K}}^{*}$
or $\boldsymbol{\mu}_{v'c'\,\nu'\boldsymbol{K}}$ with
\begin{equation}
\boldsymbol{\mu}_{vc\,\nu\boldsymbol{K}}=\int\mathrm{d}\boldsymbol{r}\,\boldsymbol{r}\,\rho_{vc\,\nu\boldsymbol{K}}\left(\boldsymbol{r}\right).
\end{equation}
The localized charge density or transition density~\cite{E2,TOE}
\begin{equation}
\rho_{vc\,\nu\boldsymbol{K}}\left(\boldsymbol{r}\right)=e\sum_{\boldsymbol{\beta}}U_{vc\,\nu\boldsymbol{K}}\left(\boldsymbol{\beta}\right)a_{c\boldsymbol{\beta}}\left(\boldsymbol{r}\right)a_{v\boldsymbol{0}}^{*}\left(\boldsymbol{r}\right),\label{eq:transcharge}
\end{equation}
with $U_{vc\,\nu\boldsymbol{K}}\left(\boldsymbol{\beta}\right)=F_{vc\,\nu}\left(\boldsymbol{\beta}\right)e^{i\alpha\boldsymbol{K}\boldsymbol{\beta}}$ is often given in terms of Wannier functions
\begin{equation}
a_{n\boldsymbol{R}}\left(\boldsymbol{r}\right)=\frac{1}{\sqrt{N}}\sum_{\boldsymbol{k}}e^{-i\boldsymbol{k}\boldsymbol{R}}\psi_{n\boldsymbol{k}}\left(\boldsymbol{r}\right)\label{eq:Wanfunc}.
\end{equation}

If $\boldsymbol{\mu}$
is parallel or perpendicular to $\boldsymbol{K}$, one speaks of longitudinal
or transversal excitons, respectively~\cite{TOE}. The nonanalytic exchange energy
therefore causes a longitudinal-transverse splitting (LT-splitting)
of spin singlet states near $K=0$. 
It is obvious that the nonanalytic exchange energy is nonzero
only for longitudinal excitons and that it is therefore connected
to a macroscopic polarization. Thus, the effect can be compared
to the LT-splitting of phonons. Since the splitting
between transverse and longitudinal excitons depends on $\left|\mu_{vc\,\nu\boldsymbol{K}}\right|^{2}$
for $vc\,\nu=v'c'\,\nu'$, it is proportional to the oscillator strength
$f_{\nu\boldsymbol{0}}$ for exciting one
exciton from the ground state of the solid by light.
This oscillator strength reads for $Ka\ll1$~\cite{TOE}
\begin{equation}
f_{\nu0}=\frac{4\delta_{S,0}}{\hbar^{2}e^{2}m_{0}}E_{\nu\boldsymbol{K}}\left|\hat{\boldsymbol{e}}_{\xi\boldsymbol{K}}\cdot\boldsymbol{\mu}_{vc\,\nu\boldsymbol{K}}\right|^{2}\label{eq:oscstr}
\end{equation}
with the energy $E_{\nu\boldsymbol{K}}$ of the exciton state~\cite{75},
the free electron mass $m_0$ and the polarization vector $\hat{\boldsymbol{e}}_{\xi\boldsymbol{K}}$
perpendicular to $\boldsymbol{K}$.
Thus, the splitting caused by $E_{\mathrm{exch}}^{\mathrm{NA}}$ 
is identical to the LT-splitting when
treating polaritons~\cite{TOE_50} and it is of appreciable size
only if the exciton is dipole allowed. 

It is now important to note that light is always transversely polarized
and that only transverse excitons are produced 
in optical absorption~\cite{E6} [cf. Eq.~(\ref{eq:oscstr})].
Longitudinal excitons cannot be seen in optical absorption spectra.
Thus, the LT-splitting in the case of polaritons increases the transverse
excitons by an energy $\Delta_{\mathrm{LT}}$. On the other hand,
the LT-splitting connected to the nonanalytic exchange interaction
increases the energy of the longitudinal excitons by the same amount
$\Delta_{\mathrm{LT}}$. Finally, both states are again degenerate
at $K=0$, which is required for reasons of symmetry. 

We can see from Eq.~(\ref{eq:HexchnaLR}) that longitudinal and transverse
exciton states are not coupled for $Ka\ll1$. As has been stated in
Ref.~\cite{TOE}, this uncoupling is ``accidental'' since it is expected
that these states are decoupled only if they transform according to
different irreducible representations of the group of $\boldsymbol{K}$~\cite{TOE_51,TOE_52}.
However,
the higher order terms $\mathcal{O}\left(K^{2}a^{2}\right)$ in Eq.~(\ref{eq:HexchnaLR})
may lead to a coupling of longitudinal and transverse exciton states
unless they transform according to different irreducible representations.
This will be shown for $\mathrm{Cu_{2}O}$
in Sec.~\ref{sub:Nonanalytic-exchange-interaction}. If a coupling
occurs, the longitudinal states will become observable in experiments
due to the admixture of transverse states~\cite{E6}.

We can now take a closer look at $m_{vc\,\nu}\left(\boldsymbol{K},\,\boldsymbol{G}\right)$
using $\boldsymbol{k}\cdot\boldsymbol{p}$ perturbation theory. It
is~\cite{SST,1}
\begin{widetext}
\begin{align}
u_{m\boldsymbol{k}}\left(\boldsymbol{r}\right) \approx &\: u_{m\boldsymbol{0}}\left(\boldsymbol{r}\right)+\frac{\hbar}{m_{0}}\sum_{n\neq m}\frac{\boldsymbol{k}\boldsymbol{p}_{nm}}{\left(E_{m}-E_{n}\right)}u_{n\boldsymbol{0}}\left(\boldsymbol{r}\right)\nonumber \\
\nonumber \\
 + & \frac{\hbar^2}{m_{0}^2}\left[\sum_{n\neq m,l\neq m}\frac{\boldsymbol{k}\boldsymbol{p}_{nl}\,\boldsymbol{k}\boldsymbol{p}_{lm}}{\left(E_{m}-E_{n}\right)\left(E_{m}-E_{l}\right)}u_{n\boldsymbol{0}}\left(\boldsymbol{r}\right)-\sum_{n\neq m,l\neq m}\frac{\boldsymbol{k}\boldsymbol{p}_{mm}\,\boldsymbol{k}\boldsymbol{p}_{nm}\delta_{nl}}{\left(E_{m}-E_{n}\right)\left(E_{m}-E_{l}\right)}u_{n\boldsymbol{0}}\left(\boldsymbol{r}\right)\right]\label{eq:kppert}
\end{align}
with $\boldsymbol{p}_{mn}=\left\langle u_{m\boldsymbol{0}}\left|\boldsymbol{p}\right|u_{n\boldsymbol{0}}\right\rangle $
and the energy $E_{n}=E_{n}\left(\boldsymbol{k}=\boldsymbol{0}\right)$ of the band
$n$ at the $\Gamma$ point. We assume that the point group of the solid contains
inversion as a group element. Then the term $\boldsymbol{p}_{mm}$ vanishes
for reasons of parity. Using the expression~(\ref{eq:kppert}), we
obtain up to second order in $\boldsymbol{K}$ and $\boldsymbol{q}$:
\begin{align}
m_{vc\,\nu}\left(\boldsymbol{K},\,\boldsymbol{G}\right) \approx &\: \frac{e}{\sqrt{N}}\sum_{\boldsymbol{q}}\, f_{vc\,\nu}\left(\boldsymbol{q}\right)\left[I_{vc}\left(\boldsymbol{G}\right)+\frac{\hbar}{m_{0}}\sum_{n\neq v}\frac{\left(\boldsymbol{q}-\gamma\boldsymbol{K}\right)\boldsymbol{p}_{vn}}{\left(E_{v}-E_{n}\right)}I_{nc}\left(\boldsymbol{G}\right)+\frac{\hbar}{m_{0}}\sum_{n\neq c}\frac{\left(\boldsymbol{q}+\alpha\boldsymbol{K}\right)\boldsymbol{p}_{nc}}{\left(E_{c}-E_{n}\right)}I_{vn}\left(\boldsymbol{G}\right)\right.\nonumber \\
\nonumber \\
  & \qquad\qquad\qquad\qquad+\frac{\hbar^2}{m_{0}^2}\sum_{n\neq v,m\neq c}\frac{\left(\boldsymbol{q}-\gamma\boldsymbol{K}\right)\boldsymbol{p}_{vn}\,\left(\boldsymbol{q}+\alpha\boldsymbol{K}\right)\boldsymbol{p}_{mc}}{\left(E_{v}-E_{n}\right)\left(E_{c}-E_{m}\right)}I_{nm}\left(\boldsymbol{G}\right)\nonumber \\
\nonumber \\
  & \qquad\qquad\qquad\qquad+\frac{\hbar^2}{m_{0}^2}\sum_{n\neq c,m\neq c}\frac{\left(\boldsymbol{q}+\alpha\boldsymbol{K}\right)\boldsymbol{p}_{nm}\,\left(\boldsymbol{q}+\alpha\boldsymbol{K}\right)\boldsymbol{p}_{mc}}{\left(E_{c}-E_{n}\right)\left(E_{c}-E_{m}\right)}I_{vn}\left(\boldsymbol{G}\right)\nonumber \\
\nonumber \\
  & \qquad\qquad\qquad\qquad+\left.\frac{\hbar^2}{m_{0}^2}\sum_{n\neq v,m\neq v}\frac{\left(\boldsymbol{q}-\gamma\boldsymbol{K}\right)\boldsymbol{p}_{mn}\,\left(\boldsymbol{q}-\gamma\boldsymbol{K}\right)\boldsymbol{p}_{vm}}{\left(E_{v}-E_{n}\right)\left(E_{v}-E_{m}\right)}I_{nc}\left(\boldsymbol{G}\right)\right].\label{eq:mKG}
\end{align}
Here we have defined $I_{mn}\left(\boldsymbol{G}\right)=\left\langle u_{m\boldsymbol{0}}\left|e^{-i\boldsymbol{G}\boldsymbol{r}}\right|u_{n\boldsymbol{0}}\right\rangle $.
The sum over $\boldsymbol{q}$ can be evaluated using
\begin{equation}
\frac{1}{\sqrt{N}}\sum_{\boldsymbol{q}}\, q_{i}^{\chi}q_{j}^{\varphi}f_{vc\,\nu}\left(\boldsymbol{q}\right)=\left.(-i)^{\chi+\varphi}\frac{\partial^{\chi}}{\partial\beta_i^{\chi}} \frac{\partial^{\varphi}}{\partial\beta_j^{\varphi}}F_{vc\,\nu}\left(\boldsymbol{\beta}\right)\right|_{\boldsymbol{\beta}=\boldsymbol{0}}\label{eq:Fgrad}
\end{equation}
with $\chi,\,\varphi=0,1,2$.

\end{widetext}
It is evident that the derivatives of the function $F_{vc\,\nu}$
at the origin must enter the exchange interaction since
we could also treat the interaction in the Wannier representation~\cite{E2}
and obtain higher order terms using a Taylor expansion at $\boldsymbol{\beta}=\boldsymbol{0}$.

Due to the special properties of the wave functions
$F_{vc\,\nu}$, the expression~(\ref{eq:Fgrad}) is nonzero only if $\varphi+\chi=L$
holds.
Therefore, we see that the leading term in Eq.~(\ref{eq:mKG})
describes the $K$ independent exchange interaction of $S$ excitons.
The terms of higher order show 
that the appearance of a $K$ dependent exchange interaction
of $S$ excitons
is inseparably connected to a $K$ \emph{in}dependent exchange
interaction of $P$ and $D$ excitons.
As the function $m_{vc\,\nu}\left(\boldsymbol{K},\,\boldsymbol{G}\right)$ enters
quadratically the exchange energy~(\ref{eq:Hexchm2}), the
relative size of the $K$ dependent exchange energy of $S$ excitons 
and the $K$ \emph{in}dependent exchange energy of $P$ excitons 
can estimated comparing 
\begin{equation}
\left|F_{vc\,\nu}\left(\boldsymbol{0}\right)\right|^2 K_0^2 = \frac{V_{\mathrm{uc}}}{\pi a_{\mathrm{exc}}^{3}}\frac{1}{n^3}K_0^2~\delta_{L,\,0}\label{eq:Sexcexch}
\end{equation}
with
\begin{equation}
\left|\left.\frac{\partial}{\partial\boldsymbol{\beta}}F_{vc\,\nu}\left(\boldsymbol{\beta}\right)\right|_{\boldsymbol{\beta}=\boldsymbol{0}}\right|^2=\frac{V_{\mathrm{uc}}}{3\pi a_{\mathrm{exc}}^{5}}\frac{n^2 -1}{n^5}~\delta_{L,\,1}.\label{eq:Pexcexch}
\end{equation}
Here we have introduced the exciton Bohr radius 
$a_{\mathrm{exc}}$ and the value $K_0$ of $K$ at the exciton photon resonance~\cite{TOE,SO}.
Note that there are always polaritons and no excitons in bulk
semiconductors due to the
coupling between excitons and photons. 
However, if this coupling is weak, it is common to speak 
of excitons and treat the interaction
within perturbation theory~\cite{TOE}.

\begin{widetext}
For the nonanalytic exchange interaction, the expression~(\ref{eq:mKG})
simplifies due to $I_{mn}\left(\boldsymbol{0}\right)=\delta_{mn}$:
\begin{align}
m_{vc\,\nu}\left(\boldsymbol{K},\,\boldsymbol{0}\right) \approx &\: \frac{e}{\sqrt{N}}\sum_{\boldsymbol{q}}\, f_{vc\,\nu}\left(\boldsymbol{q}\right)\left\{ -\frac{\hbar}{m_{0}}\frac{\boldsymbol{K}\boldsymbol{p}_{vc}}{E_{v}-E_{c}}\right.+\frac{\hbar^2}{m_{0}^2}\sum_{n\neq v,c}\left[\frac{\left(\boldsymbol{q}-\gamma\boldsymbol{K}\right)\boldsymbol{p}_{vn}\,\left(\boldsymbol{q}+\alpha\boldsymbol{K}\right)\boldsymbol{p}_{nc}}{\left(E_{v}-E_{n}\right)\left(E_{c}-E_{n}\right)}\right.\nonumber\\
  & \nonumber\\
  & \qquad+\frac{\left(\boldsymbol{q}+\alpha\boldsymbol{K}\right)\boldsymbol{p}_{vn}\,\left(\boldsymbol{q}+\alpha\boldsymbol{K}\right)\boldsymbol{p}_{nc}}{\left(E_{c}-E_{v}\right)\left(E_{c}-E_{n}\right)}+\left.\left.\frac{\left(\boldsymbol{q}-\gamma\boldsymbol{K}\right)\boldsymbol{p}_{nc}\,\left(\boldsymbol{q}-\gamma\boldsymbol{K}\right)\boldsymbol{p}_{vn}}{\left(E_{v}-E_{c}\right)\left(E_{v}-E_{n}\right)}\right]\right\} \label{eq:mvcnu}
\end{align}
It can easily be seen that $m_{vc\,\nu}\left(\boldsymbol{0},\,\boldsymbol{0}\right)=0$
holds, for which reason the nonanalytic exchange interaction does
not diverge at $K=0$. The different terms describe the nonanalytic exchange
energy of $S$ excitons ($K$ independent and $K$ dependent) and
of $P$ excitons. 
In the literature usually only the leading terms of the
exchange energy are treated, which are given by
\begin{subequations}
\begin{align}
E_{\mathrm{exch}}^{\mathrm{A}} = &\: 2\delta_{S,0}\delta_{\boldsymbol{K},\boldsymbol{K}'}\sum_{\boldsymbol{G}\neq\boldsymbol{0}}\frac{e^{2}}{\varepsilon_{0}\varepsilon V_{\mathrm{uc}}\boldsymbol{G}^{2}}F_{vc\,\nu}^{*}\left(\boldsymbol{0}\right)F_{v'c'\,\nu'}\left(\boldsymbol{0}\right)I_{vc}^{*}\left(\boldsymbol{G}\right)I_{v'c'}\left(\boldsymbol{G}\right),\label{eq:Hexcha1}
\\
\displaybreak[1]
\nonumber \\
E_{\mathrm{exch}}^{\mathrm{NA}} = &\: 2\delta_{S,0}\delta_{\boldsymbol{K},\boldsymbol{K}'}\frac{e^{2}}{\varepsilon_{0}\varepsilon V_{\mathrm{uc}}\boldsymbol{K}^{2}}F_{vc\,\nu}^{*}\left(\boldsymbol{0}\right)F_{v'c'\,\nu'}\left(\boldsymbol{0}\right)\left(\frac{\hbar}{m_{0}}\right)^{2}\frac{\boldsymbol{K}\boldsymbol{p}_{vc}^{*}\,\boldsymbol{K}\boldsymbol{p}_{v'c'}}{\left(E_{v}-E_{c}\right)\left(E_{v'}-E_{c'}\right)}.\label{eq:Hexchna1}
\end{align}
\end{subequations}
\end{widetext}
Note that
$E_{\mathrm{exch}}^{\mathrm{NA}}$ depends on $1/K^2$
and that this term cancels with the $K^2$ of the numerator.
So $E_{\mathrm{exch}}^{\mathrm{NA}}$ depends only on
the direction of $\boldsymbol{K}$ but not on its amount $K=|\boldsymbol{K}|$.
This explains the term ``nonanalytic''.

\section{Excitons in cuprous oxide~\label{sec:Excitons-in}}

Before we investigate the exchange interaction for the special case
of $\mathrm{Cu_{2}O}$, we have to discuss some specific properties
of this semiconductor. First, we have to consider the band structure
of $\mathrm{Cu_{2}O}$. Neglecting the spin-orbit coupling, the uppermost
valence band has the symmetry $\Gamma_{5}^{+}$ and is threefold degenerate
at the center of the Brillouin zone. In the literature,
this degeneracy is often accounted for by the quasi spin 
$I=1$~\cite{25,7_11,7,28,100}. This
quasi spin is a convenient abstraction to denote the three spatial
functions $\phi_{v,\, xy}$, $\phi_{v,\, yz}$ and $\phi_{v,\, zx}$,
which transform according to $\Gamma_{5}^{+}$~\cite{25,1,100}. Especially,
if we compare the states $\left|I,\, M_{I}\right\rangle $ with the
functions $\phi_{v,\, xy}$, $\phi_{v,\, yz}$ and $\phi_{v,\, zx}$
given in Ref.~\cite{8}, it is
\begin{subequations}
\begin{align}
\left|1,+1\right\rangle _{I} = &\: -\frac{1}{\sqrt{2}}\left(\phi_{v,\, yz}+i\phi_{v,zx}\right),\\
\displaybreak[1]
\nonumber \\
\left|1,0\right\rangle _{I} = &\: \phi_{v,\, xy},\\
\displaybreak[1]
\nonumber \\
\left|1,-1\right\rangle _{I} = &\: \frac{1}{\sqrt{2}}\left(\phi_{v,\, yz}-i\phi_{v,zx}\right).
\end{align}
\end{subequations}

\begin{figure}
\begin{centering}
\includegraphics[width=0.8\columnwidth]{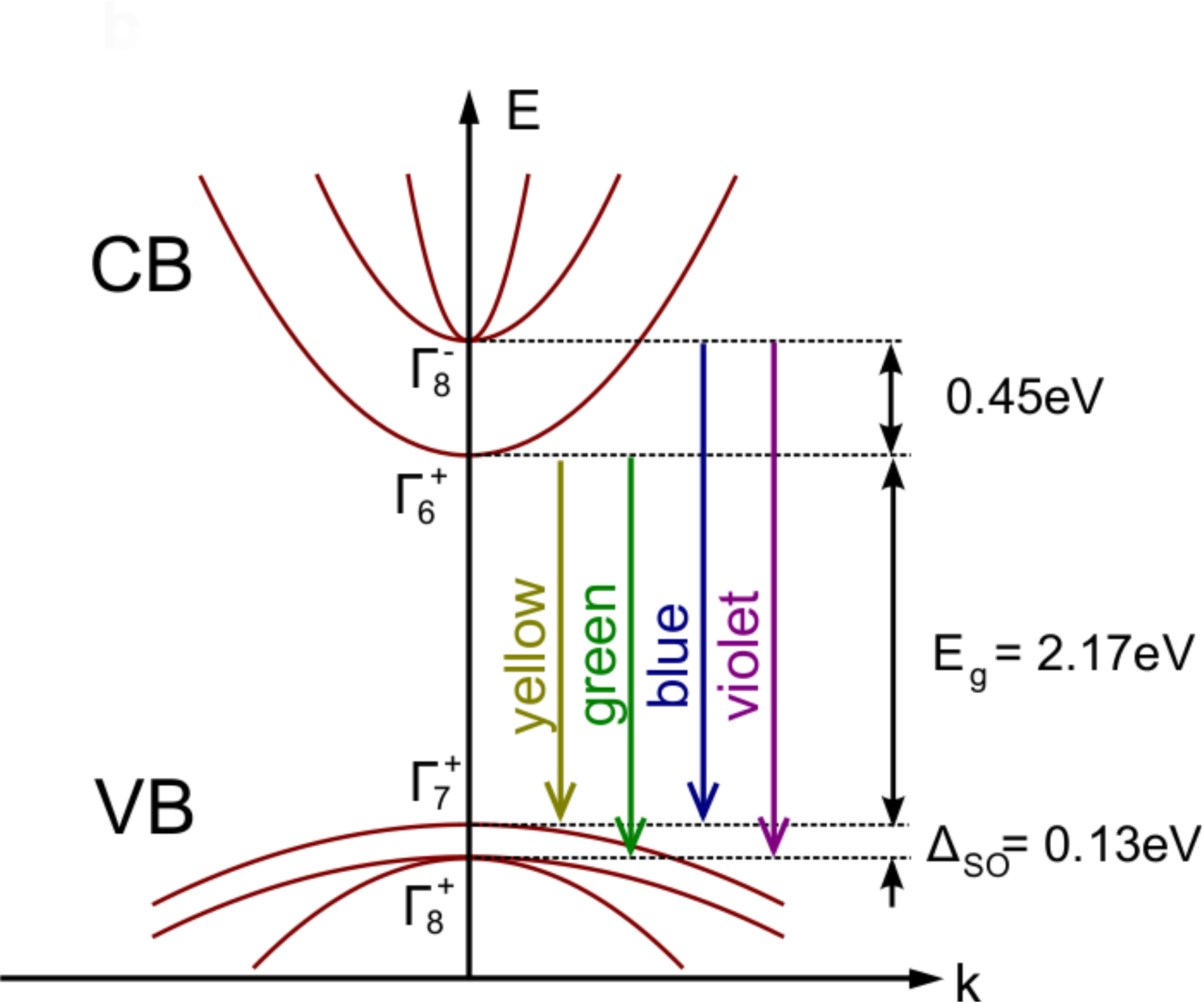}
\par\end{centering}

\protect\caption{(Color online) Band structure of $\mathrm{Cu_{2}O}$~\cite{GRE}. 
Due to the spin-orbit coupling the valence band splits into a higher lying twofold-degenerate
band $\left(\Gamma_{7}^{+}\right)$ and a lower lying fourfold-degenerate
band $\left(\Gamma_{8}^{+}\right)$. We treat the yellow and green exciton series, which 
are connected with these two valence bands and the lowest lying conduction band
of symmetry $\Gamma_{6}^{+}$.
\label{fig:Band-structure-of}}

\end{figure}

Cuprous oxide has cubic symmetry, for which reason the symmetry of
the bands is assigned by the irreducible representations $\Gamma_{i}^{\pm}$
of the cubic group $O_{\mathrm{h}}$ with the superscript $\pm$ denoting
the parity. The spin-orbit coupling between the spin $S_{\mathrm{h}}$
of a hole in the valence band and the quasi-spin $I$ splits the sixfold
degenerate band (now including the hole spin) into a higher lying
twofold-degenerate band $(\Gamma_{7}^{+})$ and a lower lying fourfold-degenerate
band $(\Gamma_{8}^{+})$ (see Fig.~\ref{fig:Band-structure-of}), which are characterized by
the effective hole spins $J=I+S_{\mathrm{h}}=1/2$ and $J=3/2$,
respectively. Within the so-called simple band model
the effective hole spin distinguishes between two independent
exciton series, i.e., the yellow $\left(J=1/2\right)$
and the green exciton series $\left(J=3/2\right)$~\cite{7,100}. 
Due to the nonspherical symmetry of the solid
and interband interactions,
the valence bands are not parabolic but deformed~\cite{100}. This
leads to a coupling between the yellow and the green exciton series,
which is described comprehensively in Ref.~\cite{100}. Here we will
discuss only the most important points.

The coupling between the valence bands or the
anisotropic dispersion 
of the orbital $\Gamma_5^+$ Bloch functions has to be considered
in the Wannier equation by the
so-called $H_{d}$-term.
The complete Hamiltonian of excitons in $\mathrm{Cu_{2}O}$ therefore reads~\cite{100}:
\begin{eqnarray}
H & = & E_{\mathrm{g}}-\frac{e^{2}}{4\pi\varepsilon_{0}\varepsilon}\frac{1}{\beta}+H_{s}\nonumber\\
\nonumber\\
 & + & H_{d}+H_{\mathrm{so}}+H_{\mathrm{exch}}+H_{\mathrm{C}}.\label{eq:Hexccu20}
\end{eqnarray}
The term 
\begin{equation}
H_{s}=\frac{\gamma'_{1}p^{2}}{2m_{0}}\label{eq:Hs}
\end{equation}
describes the average kinetic energy without the nonparabolicity and
the coupling between the bands. The $H_{d}$-term is given by
\begin{eqnarray}
H_d & = & \frac{\gamma'_{1}}{2\hbar^{2}m_{0}}\left(-\frac{\mu'}{3}P^{(2)}\cdot I^{(2)}+\frac{\sqrt{70}~\delta'}{15}\left[P^{(2)}\times I^{(2)}\right]_{0}^{(4)}\right.\nonumber\\
\nonumber\\
 & & \qquad\qquad+\left.\frac{\delta'}{3}\sum_{k=\pm4}\left[P^{(2)}\times I^{(2)}\right]_{k}^{(4)}\right)\label{eq:Hd}
\end{eqnarray}
with the irreducible tensors $P^{(2)}$ and $I^{(2)}$ defined in
Ref.~\cite{100}. The parameters $\gamma_{1}'$, $\mu'$ and $\delta'$
are connected to the Luttinger parameters of $\mathrm{Cu_{2}O}$~\cite{7,28,100}.
The $H_d$ term couples the quasi spin $I$ to the angular
momentum $L$ of the envelope function. The first summand in Eq.~(\ref{eq:Hd})
has spherical symmetry while the other terms have cubic symmetry.

The anisotropic
dispersion of the orbital $\Gamma_5^+$ Bloch functions
is in direct competition with
the spin orbit coupling
\begin{equation}
H_{\mathrm{so}}=\frac{2}{3}\Delta\left(1+\frac{1}{\hbar^{2}}\boldsymbol{I}\cdot\boldsymbol{S}_{\mathrm{h}}\right),
\end{equation}
which is diagonalized by introducing the effective hole spin $J$.
For an infinite spin orbit coupling $\Delta\rightarrow\infty$ 
the $\Gamma_7^+$ valence band would be parabolic
at the $\Gamma$ point.
However, as $\Delta=0.131\,\mathrm{eV}$~\cite{80} is 
comparatively small in $\mathrm{Cu_{2}O}$,
the nonparabolicity of 
$\Gamma_7^+$ and $\Gamma_8^+$ valence band already occurs in the 
vicinity of the $\Gamma$ point with a simultaneous mixing of both bands.

The $H_d$ term was first introduced by 
Baldereschi~\emph{et al} (see, e.g.,~Refs.~\cite{17_17_18,17_17_26,7_11,17_17,17_15}
and further references therein) to describe 
the situation for an uppermost $\Gamma_8^+$ valence band
in semiconductors like germanium mathematically correct.
The decisive breakthrough of their description
is the use of modified Bloch functions, i.e., 
Bloch functions with a lattice periodic part $u$, which
does not depend on the wave vector $k$.
These functions form a complete basis and are thus just as suitable
to describe excited states of the solid.
The modification of Baldereschi~\emph{et al} is
always required if the lattice periodic part of
the common Bloch function varies strongly with $k$.
Only due to the constant lattice periodic part the
Coulomb interaction between electron and hole
will be proportional to $1/r$ 
if the Wannier equation 
is transformed from momentum space to position 
space via a Fourier transformation
The exciton envelope function in the formalism of
Balderschi~\emph{et al} then contains
only constant $\Gamma_7^+$ and $\Gamma_8^+$ components,
i.e., the spin states with $J=1/2$ and $J=3/2$
given below.

A simple restriction to the $\Gamma_7^+$ band 
neglecting the $\Gamma_8^+$ band
and considering the nonparabolicity
via $k^4$ terms does not treat the problem correctly.
Consequently, the exchange interaction
has to be treated within the same formalism, for which reason
we use $\boldsymbol{k}\cdot\boldsymbol{p}$ perturbation theory
at the $\Gamma$ point. This is in contrast to the 
treatment by Kavoulakis~\emph{et al}~\cite{1} and to the best of our knowledge
this has not been done before.


The term $H_{\mathrm{C}}$ in Eq.~(\ref{eq:Hexccu20}) accounts for the central cell corrections~\cite{1},
which are needed to describe the $1S$ exciton correctly. Since the
radius of this exciton is very small, it is an intermediate exciton
between a Frenkel exciton and a Wannier exciton~\cite{TOE}. Therefore,
the $1S$ exciton cannot be described within the effective mass approach
due to its large extension in momentum space. However, we neglect the
central cell corrections in the following. The usage of the kinetic
energy in the form of Eqs.~(\ref{eq:Hs}) and~(\ref{eq:Hd}) and
the neglection of higher order terms in $p$ is then justified if
we use an average curvature of the bands instead of the curvature
at the center of the Brillouin zone. 
Hence, the Bohr radius of the $1S$ exciton is smaller than the one of excitons with $n\geq 2$.
Furthermore, we have to replace
the dielectric constant $\varepsilon=7.5$ by its high-frequency value
$\varepsilon_{\infty}=6.46$~\cite{1}. 

Let us consider at first the Hamiltonian~(\ref{eq:Hexccu20}) without the $H_d$ term
and the exchange interaction.
In this case we can treat spins and Wannier or Bloch functions separately from the envelope function.
The yellow and green exciton series are
described by the two states with $J=1/2$ of symmetry $\Gamma_{7}^{+}$
\begin{subequations}
\begin{align}
\left|\frac{1}{2},\,+\frac{1}{2}\right\rangle _{J} = &\: \sqrt{\frac{2}{3}}\left|+1\right\rangle _{I}\left|\downarrow\right\rangle _{\mathrm{h}}-\frac{1}{\sqrt{3}}\left|0\right\rangle _{I}\left|\uparrow\right\rangle _{\mathrm{h}},\\
\displaybreak[1]
\nonumber \\
\left|\frac{1}{2},\,-\frac{1}{2}\right\rangle _{J} = &\: \frac{1}{\sqrt{3}}\left|0\right\rangle _{I}\left|\downarrow\right\rangle _{\mathrm{h}}-\sqrt{\frac{2}{3}}\left|-1\right\rangle _{I}\left|\uparrow\right\rangle _{\mathrm{h}}.
\end{align}
\end{subequations}
and the four states with $J=3/2$ of symmetry $\Gamma_{8}^{+}$
\begin{subequations}
\begin{align}
\left|\frac{3}{2},\,+\frac{3}{2}\right\rangle _{J} = &\: \left|+1\right\rangle _{I}\left|\uparrow\right\rangle _{\mathrm{h}},\\
\displaybreak[1]
\nonumber \\
\left|\frac{3}{2},\,+\frac{1}{2}\right\rangle _{J} = &\: \frac{1}{\sqrt{3}}\left|1\right\rangle _{I}\left|\downarrow\right\rangle _{\mathrm{h}}+\sqrt{\frac{2}{3}}\left|0\right\rangle _{I}\left|\uparrow\right\rangle _{\mathrm{h}}.\\
\displaybreak[1]
\nonumber \\
\left|\frac{3}{2},\,-\frac{1}{2}\right\rangle _{J} = &\: \sqrt{\frac{2}{3}}\left|0\right\rangle _{I}\left|\downarrow\right\rangle _{\mathrm{h}}+\frac{1}{\sqrt{3}}\left|-1\right\rangle _{I}\left|\uparrow\right\rangle _{\mathrm{h}},\\
\displaybreak[1]
\nonumber \\
\left|\frac{3}{2},\,-\frac{3}{2}\right\rangle _{J} = &\: \left|-1\right\rangle _{I}\left|\downarrow\right\rangle _{\mathrm{h}}.
\end{align}
\end{subequations}
If we now add the
electron spin and the Wannier function $\phi_{c,\, s}$ of the conduction
band, which transform together according to $\Gamma_{6}^{+}\otimes\Gamma_{1}^{+}=\Gamma_{6}^{+}$,
we obtain states with the total momentum $G=J+S_{\mathrm{e}}=0$ and $G=1$.
Subsequently, these states have to be multiplied by the
hydrogen-like envelope function $F_{\nu}\left(\boldsymbol{\beta}\right)$.

In the Cartesian basis the ground states of the yellow exciton are (cf. Ref.~\cite{8})
\begin{subequations}
\begin{align}
\left|P\right\rangle = &\:  F_{1,0,0}\left(\boldsymbol{\beta}\right)\left|0,\,0\right\rangle _{G},
\\
\displaybreak[1]
\nonumber \\
\left|O_{xy}\right\rangle = &\:  F_{1,0,0}\left(\boldsymbol{\beta}\right)\left|1,\,0\right\rangle _{G},
\\
\displaybreak[1]
\nonumber \\
\left|O_{yz}\right\rangle = &\:  \frac{1}{\sqrt{2}}F_{1,0,0}\left(\boldsymbol{\beta}\right)\left(\left|1,\,-1\right\rangle _{G}-\left|1,\,+1\right\rangle _{G}\right),
\\
\displaybreak[1]
\nonumber \\
\left|O_{zx}\right\rangle = &\:  \frac{i}{\sqrt{2}}F_{1,0,0}\left(\boldsymbol{\beta}\right)\left(\left|1,\,-1\right\rangle _{G}+\left|1,\,+1\right\rangle _{G}\right).
\end{align}
\end{subequations}
The state $\left|P\right\rangle$ of symmetry $\Gamma_2^+$ 
is the para exciton state and the states $|O_{ij}\rangle$ of symmetry $\Gamma_5^+$ 
are the ortho exciton states. It is possible to express these
states using the eigenstates of the spin $S=S_{\mathrm{e}}+S_{\mathrm{h}}$~\cite{8}:
\begin{widetext}
\begin{subequations}
\begin{align}
\left|P\right\rangle  = &\: \frac{1}{\sqrt{6}}F_{1,0,0}\left(\boldsymbol{\beta}\right)\phi_{c,\, s}\left[\sqrt{2}\phi_{v,\, xy}\left|1,\,0\right\rangle _{S}+\left(-\phi_{v,\, yz}+i\phi_{v,\, zx}\right)\left|1,\,+1\right\rangle _{S}+\left(\phi_{v,\, yz}+i\phi_{v,\, zx}\right)\left|1,\,-1\right\rangle _{S}\right],\\
\displaybreak[1]
\nonumber \\
\left|O_{xy}\right\rangle  = &\: \frac{-1}{\sqrt{6}}F_{1,0,0}\left(\boldsymbol{\beta}\right)\phi_{c,\, s}\left[-\sqrt{2}\phi_{v,\, xy}\left|0,\,0\right\rangle _{S}+\left(\phi_{v,\, yz}-i\phi_{v,\, zx}\right)\left|1,\,+1\right\rangle _{S}+\left(\phi_{v,\, yz}+i\phi_{v,\, zx}\right)\left|1,\,-1\right\rangle _{S}\right],\\
\displaybreak[1]
\nonumber \\
\left|O_{yz}\right\rangle  = &\: \frac{1}{\sqrt{6}}F_{1,0,0}\left(\boldsymbol{\beta}\right)\phi_{c,\, s}\left[\phi_{v,\, xy}\left|1,\,+1\right\rangle _{S}+\phi_{v,\, xy}\left|1,\,-1\right\rangle _{S}+i\sqrt{2}\phi_{v,\, yz}\left|0,\,0\right\rangle _{S}+\sqrt{2}\phi_{v,\, zx}\left|1,\,0\right\rangle _{S}\right],\\
\displaybreak[1]
\nonumber \\
\left|O_{zx}\right\rangle  = &\: \frac{-i}{\sqrt{6}}F_{1,0,0}\left(\boldsymbol{\beta}\right)\phi_{c,\, s}\left[\phi_{v,\, xy}\left|1,\,+1\right\rangle _{S}-\phi_{v,\, xy}\left|1,\,-1\right\rangle _{S}+\sqrt{2}\phi_{v,\, yz}\left|1,\,0\right\rangle _{S}+i\sqrt{2}\phi_{v,\, zx}\left|0,\,0\right\rangle _{S}\right].
\end{align}
\label{eq:deltainf}
\end{subequations}
\end{widetext}
One can see that the para exciton state does not contain a singlet
component, i.e., a component with $S=0$. Therefore, this state is
spin-flip forbidden in optical excitations, which explains the term
``para'' or ``dark'' exciton~\cite{GRE}. However, we may note
at this point that the ortho and para exciton states are not eigenstates
of the operators $S^{2}$ and $S_{z}$. Therefore, it may be misleading
to speak of singlet and triplet states~\cite{GRE,9_1}. 

The exciton states are generally mixed by the $H_{d}$ term~(\ref{eq:Hd}) due to the
coupling between $L$ and $I$. 
Since parity is a good quantum number in $\mathrm{Cu_{2}O}$, the
$H_d$ term mixes only exciton states with even values of $L$
or with odd values of $L$~\cite{7}.
Therefore, $D$ excitons are admixed to $S$ excitons and vice versa.
The coupling due to the $H_{d}$ term leads to an energy gain in the system, 
which was discussed in Ref.~\cite{100}.

As the radius of the yellow $1S$ exciton is small in position space,
it is extended in momentum space, for which reason
we expect its coupling to the green series to be strong.
Due to the admixture, the yellow ortho exciton becomes more and more a
pure singlet state as the total spin $S=S_{\mathrm{e}}+S_{\mathrm{h}}$
is a good quantum number in the limit of $\Delta=0$.

In this limiting case with $\Delta=0$, the introduction of the 
effective hole spin $J$ would not be
necessary. The exciton wave function could be written as the product
of a space function, which also depends on $I$, and a spin function.
Without the $H_d$ term the ground states of the exciton would then read
\begin{subequations}
\begin{align}
\left|P_{1,j}\right\rangle  = &\: F_{1,0,0}\left(\boldsymbol{\beta}\right)\phi_{c,\, s}\phi_{v,\, j}\left|1,\,1\right\rangle _{S},\\
\displaybreak[1]
\nonumber \\
\left|P_{0,j}\right\rangle  = &\: F_{1,0,0}\left(\boldsymbol{\beta}\right)\phi_{c,\, s}\phi_{v,\, j}\left|1,\,0\right\rangle _{S},\\
\displaybreak[1]
\nonumber \\
\left|P_{-1,j}\right\rangle  = &\: F_{1,0,0}\left(\boldsymbol{\beta}\right)\phi_{c,\, s}\phi_{v,\, j}\left|1,\,-1\right\rangle _{S},\\
\displaybreak[1]
\nonumber \\
\left|O_{j}\right\rangle  = &\: F_{1,0,0}\left(\boldsymbol{\beta}\right)\phi_{c,\, s}\phi_{v,\, j}\left|0,\,0\right\rangle _{S},
\end{align}
\label{eq:delta0}%
\end{subequations}
with $j=xy,\, yz,\, zx$. In this case there are also three ortho exciton states.
The para and ortho exciton states are true triplet states $\left(S=1\right)$ and singlet
states $\left(S=0\right)$, respectively.

The Hamiltonian~(\ref{eq:Hexccu20}) is given for $K=0$.
In the general case with $K\neq0$ additional terms appear~\cite{100}:
\begin{align}
T_{\mathrm{t}}\left(\boldsymbol{K}\right) = &\: \Omega_{1}K^{2}\boldsymbol{1}\nonumber\\
 - & \Omega_{3}\left(K_{1}^{2}\left(3\boldsymbol{I}_{1}^{2}-2\hbar^{2}\boldsymbol{1}\right)+\mathrm{c.p.}\right)/\hbar^{2}\nonumber\\
 - & \Omega_{5}\left(K_{1}K_{2}\left(\boldsymbol{I}_{1}\boldsymbol{I}_{2}+\boldsymbol{I}_{2}\boldsymbol{I}_{1}\right)+\mathrm{c.p.}\right)/\hbar^{2}\label{eq:TtK}
\end{align}
As can be seen, these $K^{2}$ dependent terms are $3\times3$ matrices
and can again be divided into an $H_{s}$-term, an $H_{d}$-term of
spherical symmetry and an $H_{d}$-term of cubic symmetry, i.e.,
we can write
\begin{align}
T_{\mathrm{t}}\left(\boldsymbol{K}\right) = &\: \left(\Omega_1\right) K^2-\left(\frac{\Omega_5+2\Omega_3}{15\hbar^2}\right)K^{(2)}\cdot I^{(2)}\nonumber\\
+ & \left(\frac{\Omega_5-3\Omega_3}{18\hbar^2}\right)\left(\frac{\sqrt{70}}{5}\left[K^{(2)}\times I^{(2)}\right]_{0}^{(4)}\right.\nonumber\\
& \qquad\qquad\quad+\left.\sum_{k=\pm4}\left[K^{(2)}\times I^{(2)}\right]_{k}^{(4)}\right).
\end{align}
To describe
the exciton series in $\mathrm{Cu_{2}O}$ correctly, the Schr\"odinger
equation with the operators~(\ref{eq:Hexccu20}) and~(\ref{eq:TtK})
has to be solved for fixed values of $K$. However, as the effect
of the $K^{2}$ dependent terms on the relative motion is small, 
the effect of $T_{\mathrm{t}}\left(\boldsymbol{K}\right)$
can be treated within order perturbation theory.

It has been shown in Ref.~\cite{100}
that the coefficients $\Omega$ in $T_{\mathrm{t}}\left(\boldsymbol{K}\right)$
are of the correct order of magnitude to describe
the $K$ dependent splitting of the $1S$ exciton state, which was observed
experimentally and originally assigned to the exchange interaction~\cite{9_1,8,9}.
In the next section~\ref{sec:Exchange-interaction-for}
we will show that the exchange interaction is far
too small to explain this splitting.

\section{Exchange interaction for cuprous oxide~\label{sec:Exchange-interaction-for}}

In this section we want to estimate the maximum size of the exchange interaction for
the exciton ground state in $\mathrm{Cu_{2}O}$ following the explanations given in Refs.~\cite{8,1}. 
Note that it would be necessary to 
solve the full exciton Hamiltonian~(\ref{eq:Hexccu20}) including all $K$ dependent
terms to determine the true size of the exchange interaction.
As has been stated in Sec.~\ref{sec:Excitons-in}, parity is a good quantum number and the
exciton ground state contains mainly $S$ like but also $D$ like envelope functions.
Due to the results of Sec.~\ref{sec:Exchange-interaction} the 
($K$ dependent and $K$ independent) exchange interaction is
strongest if the envelope function is purely $S$ like
and if $\nu=1$ holds.
Furthermore, for the
exchange interaction only the singlet component of the states is of
importance. From Eqs.~(\ref{eq:deltainf}) and~(\ref{eq:delta0})
we see that we can set
\begin{subequations}
\begin{align}
\rho_{vc\,\nu\boldsymbol{K}}^{(P)}\left(\boldsymbol{r}\right) = &\: 0,\\
\displaybreak[1]
\nonumber \\
\rho_{vc\,\nu\boldsymbol{K}}^{(O)}\left(\boldsymbol{r}\right) = &\: c_{\rho}e\sum_{\boldsymbol{\beta}}U_{vc\,\nu\boldsymbol{K}}\left(\boldsymbol{\beta}\right)\nonumber\\
  & \qquad\qquad\times\phi_{c,\, s}\left(\boldsymbol{r}-\boldsymbol{\beta}\right)\phi_{v,\, j}^{*}\left(\boldsymbol{r}\right),\label{eq:rhoortho}
\end{align}
\end{subequations}
with $j=xy,\, yz,\, zx$. 
The pre\-factor $c_{\rho}$
is of the order $1$. Even though the exchange energy is not diagonal
with respect to $\nu$, we consider only the dominant contribution
$E_{\mathrm{exch}}\left(vc\,1S\,\boldsymbol{K},\, v'c\,1S\,\boldsymbol{K}'\right)$
with $\nu=\nu'=1S$ or more precisely $\nu=(n,L,M)=(1,0,0)$.

\subsection{Analytic exchange interaction~\label{sub:Analytic-exchange-interaction}}

The $K$ dependence of the analytic exchange interaction has been
neglected in Ref.~\cite{1} and will be treated here. We
estimate its magnitude to show that the $K$ dependent splitting of the $1S$
exciton state treated in Refs.~\cite{9_1,8,9} cannot be explained
in terms of the exchange interaction.

In the case of the analytic exchange interaction we consider only the zero and first order
terms in the function $m_{vc\,\nu}\left(\boldsymbol{K},\,\boldsymbol{G}\right)$ of Eq.~(\ref{eq:mKG}).
As can be seen from Eq.~(\ref{eq:Hexchm2}), the analytic exchange energy depends on 
$m_{vc\,\nu}^{*}\left(\boldsymbol{K},\,\boldsymbol{G}\right)m_{v'c'\,\nu'}\left(\boldsymbol{K},\,\boldsymbol{G}\right)$.
When calculating the exchange energy the second order terms in 
$m_{v'c'\,\nu'}\left(\boldsymbol{K},\,\boldsymbol{G}\right)$ have to be multiplied with the zero order term of
$m_{vc\,\nu}^{*}\left(\boldsymbol{K},\,\boldsymbol{G}\right)$ and vice versa. Since 
the zero order term is a diagonal $3\times3$ matrix, the resulting $K^2$ dependent 
terms cannot describe a $K$ dependent splitting of the exciton ground state.
Furthermore, we will estimate the size of these terms in the 
following and show that they are negligibly small.

We can write
\begin{widetext}
\begin{align}
m_{vc\,\nu}\left(\boldsymbol{K},\,\boldsymbol{G}\right) \approx &\: \frac{ec_{\rho}}{\sqrt{N}}\sum_{\boldsymbol{q}}\, f_{vc\,\nu}\left(\boldsymbol{q}\right)\left[I_{vc}\left(\boldsymbol{G}\right)+\frac{\hbar}{m_{0}}\left\{ \sum_{n\neq v}\frac{\left(\boldsymbol{q}-\gamma\boldsymbol{K}\right)\boldsymbol{p}_{vn}}{\left(E_{v}-E_{n}\right)}I_{nc}\left(\boldsymbol{G}\right)+\sum_{n\neq c}\frac{\left(\boldsymbol{q}+\alpha\boldsymbol{K}\right)\boldsymbol{p}_{nc}}{\left(E_{c}-E_{n}\right)}I_{vn}\left(\boldsymbol{G}\right)\right\} \right]\nonumber \\
\nonumber \\
 = & ec_{\rho}F_{vc\,\nu}\left(\boldsymbol{0}\right)I_{vc}\left(\boldsymbol{G}\right)\nonumber\\
 \nonumber\\
 + & ec_{\rho}\frac{\hbar}{m_{0}}\left[\left(-i\nabla_{\boldsymbol{\beta}}\right)F_{vc\,\nu}\left(\boldsymbol{\beta}\right)\right]_{\boldsymbol{\beta}=\boldsymbol{0}}\left\{\sum_{n\neq v}\frac{\boldsymbol{p}_{vn}I_{nc}\left(\boldsymbol{G}\right)}{\left(E_{v}-E_{n}\right)}+\sum_{n\neq c}\frac{\boldsymbol{p}_{nc}I_{vn}\left(\boldsymbol{G}\right)}{\left(E_{c}-E_{n}\right)}\right\} \nonumber \\
\nonumber \\
 + & ec_{\rho}\frac{\hbar}{m_{0}}F_{vc\,\nu}\left(\boldsymbol{0}\right)\left\{\sum_{n\neq v}\frac{-\gamma\boldsymbol{K}\boldsymbol{p}_{vn}}{\left(E_{v}-E_{n}\right)}I_{nc}\left(\boldsymbol{G}\right)+\sum_{n\neq c}\frac{\alpha\boldsymbol{K}\boldsymbol{p}_{nc}}{\left(E_{c}-E_{n}\right)}I_{vn}\left(\boldsymbol{G}\right)\right\}.\label{eq:mvc_beta}
\end{align}
If we now set $\nu=\nu'=1S$, the gradient of $F_{vc\,1S}\left(\boldsymbol{\beta}\right)$
at $\boldsymbol{\beta}=\boldsymbol{0}$ vanishes. Finally, we have
\begin{align}
E_{\mathrm{exch}}^{\mathrm{A}} = &\: 2\delta_{S,0}\delta_{\boldsymbol{K},\boldsymbol{K}'}\frac{e^{2}c_{\rho}^{2}}{\varepsilon_{0}\varepsilon_{\infty}\pi a_{\mathrm{exc}}^{3}}\sum_{\boldsymbol{G}\neq\boldsymbol{0}}\frac{1}{\left(\boldsymbol{K}+\boldsymbol{G}\right)^{2}}\nonumber \\
\nonumber \\
 \times & \left[I_{v'c}\left(\boldsymbol{G}\right)+\frac{\hbar}{m_{0}}\left\{ \sum_{n\neq v'}\frac{-\gamma\boldsymbol{K}\boldsymbol{p}_{v'n}}{\left(E_{v'}-E_{n}\right)}I_{nc}\left(\boldsymbol{G}\right)+\sum_{n\neq c}\frac{\alpha\boldsymbol{K}\boldsymbol{p}_{nc}}{\left(E_{c}-E_{n}\right)}I_{v'n}\left(\boldsymbol{G}\right)\right\} \right]\nonumber \\
\nonumber \\
 \times & \left[I_{vc}\left(\boldsymbol{G}\right)+\frac{\hbar}{m_{0}}\left\{ \sum_{n\neq v}\frac{-\gamma\boldsymbol{K}\boldsymbol{p}_{vn}}{\left(E_{v}-E_{n}\right)}I_{nc}\left(\boldsymbol{G}\right)+\sum_{n\neq c}\frac{\alpha\boldsymbol{K}\boldsymbol{p}_{nc}}{\left(E_{c}-E_{n}\right)}I_{vn}\left(\boldsymbol{G}\right)\right\} \right]^{*}.\label{eq:Hexcha1S}
\end{align}
\end{widetext}
The component with $\boldsymbol{K}=\boldsymbol{0}$ describes the
experimentally observed splitting between ortho and para excitons
of $12\,\mathrm{meV}$~\cite{1_6a,E2_16,1_6c}.
Therefore, we set
\begin{equation}
12\,\mathrm{meV}=\frac{2e^{2}c_{\rho}^{2}}{\varepsilon_{0}\varepsilon_{\infty}\pi a_{\mathrm{exc}}^{3}}\sum_{\boldsymbol{G}\neq\boldsymbol{0}}\frac{1}{\boldsymbol{G}^{2}}\left|I_{vc}\left(\boldsymbol{G}\right)\right|^{2}.\label{eq:12meV}
\end{equation}
A restriction to the six summands with the smallest
value $\boldsymbol{G}_{0}$ of $\boldsymbol{G}$ 
as in Ref.~\cite{1} is in general not correct.
Due to the symmetry of the Bloch functions
other values of $\boldsymbol{G}$ will contribute
even more strongly to the sum in Eq.~(\ref{eq:Hexcha1S}).
Indeed, it is worth mentioning that the
symmetry group of the lattice in
$\mathrm{Cu_{2}O}$ is only isomorphic to the cubic group
$O_{\mathrm{h}}$~\cite{SO}. Since the Cu atoms in $\mathrm{Cu_{2}O}$
form an fcc sublattice, it can be seen from the unit cell of
$\mathrm{Cu_{2}O}$ that the lattice is not invariant under
reflections but under a glide reflection with a
translation of $a/2$ (see also supplementary material of Ref.~\cite{GRE}), where $a$ denotes the lattice constant
$a=4.26\times10^{-10}\,\mathrm{m}$ of $\mathrm{Cu_{2}O}$~\cite{JPCS27,P6,DB_45}.
The Bloch functions must be invariant under
this operation. If we write
$u_{n\boldsymbol{K}}\left(\boldsymbol{r}\right)=
\sum_{\boldsymbol{G}}C_{n\boldsymbol{K}}\left(\boldsymbol{G}\right)e^{i\boldsymbol{G}\boldsymbol{r}}$~\cite{SST},
we see that the vector components of $\boldsymbol{G}$ can only take whole-number multiples of
$4\pi/a$ instead of $2\pi/a$.

The $K$ dependence of the analytic exchange interaction arises from
the $\boldsymbol{K}\boldsymbol{p}_{mn}$-terms and
the factor $1/\left(\boldsymbol{K}+\boldsymbol{G}\right)^{2}$ 
in Eq.~(\ref{eq:Hexcha1S}).
At first, we will estimate the effect of the $\boldsymbol{K}\boldsymbol{p}_{mn}$-terms.
Due to reasons of symmetry, the terms linear in $K$
must vanish when evaluating the product in Eq.~(\ref{eq:Hexcha1S}).
The $K^{2}$ dependent terms are of the same order of magnitude as the
second order terms in the function $m_{vc\,\nu}\left(\boldsymbol{K},\,\boldsymbol{G}\right)$, which we have neglected.
We can now use Eq.~(\ref{eq:12meV}) to give an upper limit for their magnitude
and to prove that their neglection is justified.
Using the values $\left|\boldsymbol{p}_{nm}\right|/\hbar\approx 1.3\times10^{9}\,\mathrm{m}^{-1}$
and $\left(E_{m}-E_{n}\right)\geq\Delta E=449\,\mathrm{meV}$ given
in Ref.~\cite{1}, we obtain
\begin{equation}
\left(\frac{\hbar}{m_{0}}\left|\boldsymbol{p}_{nm}\right|K_{0}\frac{1}{\Delta E}\right)^{2}\times12\,\mathrm{meV}\approx0.4\,\mathrm{\mu eV}.\label{eq:absch1}
\end{equation}
We see that this part of the analytic exchange interaction is very small. 

However, we have shown in Sec.~\ref{sec:Exchange-interaction} that a $K$ dependent
exchange interaction of $S$ excitons is connected to 
a $K$ independent analytic exchange interaction of $P$ excitons.
Using the
result of Eq.~(\ref{eq:absch1}), we can estimate the size of the analytic exchange energy
of the $2P$ excitons via Eqs.~(\ref{eq:Sexcexch}) and~(\ref{eq:Pexcexch}).
With the exciton Bohr radius $a_{\mathrm{exc}}=0.53~\mathrm{nm}$
of the $1S$ exciton, the corresponding value $a_{\mathrm{exc}}=1.1~\mathrm{nm}$
for $P$ excitons~\cite{1} and $K_{0}=2.62\times10^{7}\,\mathrm{m}^{-1}$~\cite{8},
the maximum size of the analytic exchange energy
of the $2P$ excitons is
\begin{equation}
\frac{(0.53)^3}{3 (1.1)^5 (0.0262)^2}\frac{2^2-1}{2^5}\times 0.4~\mathrm{\mu eV}\approx 1.6~\mathrm{\mu eV}.
\end{equation}
We see that also this energy
is negligibly small. Furthermore, the line widths of the $P$ excitons
in $\mathrm{Cu_{2}O}$ are too large to detect a splitting in the
order of a few $\mathrm{\mu eV}$.

Let us now treat the $K$ dependence arising from the prefactor $1/\left(\boldsymbol{K}+\boldsymbol{G}\right)^{2}$.
This factor can written as a Fourier series at $\boldsymbol{K}=\boldsymbol{0}$ for $\boldsymbol{K}\ll\boldsymbol{G}$,
\begin{align}
\frac{1}{\left(\boldsymbol{K}+\boldsymbol{G}\right)^{2}} \approx &\: \frac{1}{G^{2}}-\frac{2\left(\boldsymbol{K}\boldsymbol{G}\right)}{G^{4}}\nonumber\\
+ & \frac{1}{G^{6}}\boldsymbol{G}^{T}\left[-\boldsymbol{1}K^{2}+4\boldsymbol{K}\cdot\boldsymbol{K}^{\mathrm{T}}\right]\boldsymbol{G}.
\end{align}
Inserting this expression
in Eq.~(\ref{eq:Hexcha1S}), the term proportional to $K$ vanishes
for reasons of symmetry. The magnitude of the $K^{2}$ dependent term can
be estimated assuming that $K$ is oriented in $\left[100\right]$
direction and using the reciprocal lattice vectors with the smallest modulus $4\pi/a$. This gives an upper limit
of
\begin{equation}
12\,\mathrm{meV}\times3 K_{0}^{2}\left(\frac{a}{4\pi}\right)^{2}\approx28\,\mathrm{neV}\label{eq:absch2}
\end{equation}
for the prefactor of those $K^{2}$ dependent terms 
in Eq.~(\ref{eq:Hexcha1S})
which originate from the Fourier expansion of 
$1/\left(\boldsymbol{K}+\boldsymbol{G}\right)^{2}$.
We see that not only the result of Eq.~(\ref{eq:absch1}) but also
the result of Eq.~(\ref{eq:absch2}) is one magnitude smaller than
the experimentally observed values for the $K$ dependent splitting
of the $1S$ exciton~\cite{8}. As the estimated values are upper
limits for the prefactors, the actual magnitude of the analytic exchange
interaction is generally much smaller.

However, using group theoretical considerations, it is obvious that in both cases
the $K^{2}$ dependent
terms can be written as a sum of the invariant matrices $\boldsymbol{1}K^{2}$,
$(3K_{i}^{2}-K^{2})\left(\hat{\boldsymbol{e}}_{i}\otimes\hat{\boldsymbol{e}}_{i}\right)$
and $K_{i}K_{j}\left(\hat{\boldsymbol{e}}_{i}\otimes\hat{\boldsymbol{e}}_{j}\right)$
with $i,j=1,2,3$ and $i\neq j$, since every $K$ dependent energy
as regards states of the symmetry $\Gamma_{5}^{+}$ must lead to matrices
of this form~\cite{8,G3}. This can be seen, e.g., from Eq.~(\ref{eq:TtK}),
where the dispersion of the exciton is described by the same matrices.
Hence, the $K$ dependent splitting of the $1S$ exciton states must
in any case be described by matrices of this form~\cite{8}. However,
from the experimental point of view the physical origin of these matrices
is \emph{a priori} unknown. In Refs.~\cite{9_1,8,9} it has been assumed
that the exchange interaction is responsible for the $K$ dependent
splitting.

We have now shown that the $K$ dependent analytic
exchange interaction is negligibly small in $\mathrm{Cu_{2}O}$ and
that it cannot explain the $K$ dependent splitting of the $1S$ exciton.
Furthermore, due to the specific form of the exchange interaction, 
it would not be experimentally distinguishable
from the dispersion of the exciton described by Eq.~(\ref{eq:TtK}).

Only the $K$ dependent nonanalytic
exchange interaction may contribute to the splitting of the $1S$ ortho exciton. This will be
investigated in the following section~\ref{sub:Nonanalytic-exchange-interaction}.

\subsection{Nonanalytic exchange interaction~\label{sub:Nonanalytic-exchange-interaction}}

We will now treat the nonanalytic exchange interaction for $\mathrm{Cu_{2}O}$. 
As the conduction
band and the valence band in $\mathrm{Cu_{2}O}$ have the same (positive)
parity and the momentum operator $\boldsymbol{p}$ has negative parity,
the matrix element $\boldsymbol{p}_{vc}=\left\langle u_{v\boldsymbol{0}}\left|\boldsymbol{p}\right|u_{c\boldsymbol{0}}\right\rangle $
vanishes. Therefore, the main contribution to the nonanalytic exchange
interaction comes from the term in square
brackets in Eq.~(\ref{eq:mvcnu}). 

We can see again the close connection between the nonanalytic exchange
interaction and the oscillator strength: Inserting the $q\, K$ dependent
terms of Eq.~(\ref{eq:mvcnu}) into Eq.~(\ref{eq:Hexchm2}), one
obtains the $K$ independent nonanalytic exchange energy of $P$ excitons.
Since these excitons are dipole-allowed, their oscillator strength
is also $K$ independent. The exchange energy exactly equals the LT-splitting
when treating $P$ exciton-polaritons. 

The $K^{2}$ dependent terms of Eq.~(\ref{eq:mvcnu}) will lead to
a $K^{2}$ dependent exchange energy for the $S$ excitons. These
excitons are quadrupole allowed and their oscillator
strength is also $K^{2}$ dependent. For reasons of symmetry, the
energy difference between longitudinal and transversal $S$ excitons
at $K=0$ is exactly zero, as well. The fact that $S$ excitons are
quadrupole allowed for finite values of $K$ can be understood from
a symmetry reduction: The cubic group reduces for finite values of
$K$ to a group of lower symmetry, e.g., $C_{4v}$, which does not
contain inversion as a group element~\cite{G3,TOE_51}. This leads
to a $K$ dependent admixture of $P$ excitons to $S$ excitons.

In the following, we will concentrate on the $K^{2}$ dependent 
exchange energy of the $1S$ excitons to estimate its magnitude 
and investigate its angle dependency.
Due to the close connection between exchange energy
and oscillator strength, we expect the ratio of the $K^{2}$ dependent exchange energy of $S$ excitons
and the $K$ independent exchange energy of $P$ excitons to be of the same size
as the ratio of the corresponding oscillator strengths.

We can write
\begin{widetext}
\begin{equation}
m_{vc\,1S}\left(\boldsymbol{K},\,\boldsymbol{0}\right) \approx \frac{ec_{\rho}\hbar^{2}}{m_{0}^{2}}F_{vc\,1S}\left(\boldsymbol{0}\right)\left(\left\langle u_{v\boldsymbol{0}}\right|\left(\boldsymbol{K}\boldsymbol{p}\right)\left[\sum_{n\neq v,c}g_{vc}\left(E_{n}\right)\left|u_{n\boldsymbol{0}}\right\rangle \left\langle u_{n\boldsymbol{0}}\right|\right]\left(\boldsymbol{K}\boldsymbol{p}\right)\left|u_{c\boldsymbol{0}}\right\rangle\right) \label{eq:KuuK}
\end{equation}
with
\begin{equation}
g_{vc}\left(E_{n}\right)=\frac{\gamma\alpha\left(E_{v}-E_{c}\right)-\alpha^{2}\left(E_{v}-E_{n}\right)+\gamma^{2}\left(E_{c}-E_{n}\right)}{\left(E_{v}-E_{n}\right)\left(E_{c}-E_{n}\right)\left(E_{v}-E_{c}\right)}.
\end{equation}
\end{widetext}
Using group theory, we can determine the non-vanishing terms of the
exchange energy. The operator in square brackets in Eq.~(\ref{eq:KuuK})
is a projection operator. For reasons of symmetry this operator has
to transform according to the irreducible representation $\Gamma_{1}^{+}$.
On the other hand, the operator $\boldsymbol{p}$ transforms according
to $\Gamma_{4}^{-}$. The symmetry of the operator between the Bloch
functions is therefore
\begin{equation}
\Gamma_{4}^{-}\otimes\Gamma_{1}^{+}\otimes\Gamma_{4}^{-}=\Gamma_{1}^{+}\oplus\Gamma_{3}^{+}\oplus\Gamma_{4}^{+}\oplus\Gamma_{5}^{+}.
\end{equation}
The symmetry of the Bloch functions is 
\begin{equation}
\Gamma_{5}^{+}\otimes\Gamma_{1}^{+}=\Gamma_{5}^{+}.
\end{equation}
Consequently, the expression~(\ref{eq:KuuK}) does not vanish only
if the operator has the symmetry $\Gamma_{5}^{+}$~\cite{G1}. We
can then consider the coupling coefficients for the case $\Gamma_{4}^{-}\otimes\Gamma_{4}^{-}\rightarrow\Gamma_{5}^{+}$.
With the basis functions $\left|X\right\rangle $, $\left|Y\right\rangle $,
$\left|Z\right\rangle $ of $\Gamma_{4}^{-}$ and the basis functions
$|\tilde{X}\rangle=\left|YZ\right\rangle $,
$|\tilde{Y}\rangle=\left|ZX\right\rangle $, and $|\tilde{Z}\rangle=\left|XY\right\rangle $
of $\Gamma_{5}^{+}$, we see that, e.g., the $\Gamma_{5}^{+}$ like part of the products
$\left|X\right\rangle _{1}\left|Y\right\rangle _{2}$ and $\left|Y\right\rangle _{1}\left|X\right\rangle _{2}$
transforms as $|\tilde{Z}\rangle /\sqrt{2}$~\cite{G3}. So we write
\begin{equation}
\langle\tilde{Z}|\left(\left|X\right\rangle _{1}\left|Y\right\rangle _{2}\right)=\frac{1}{\sqrt{2}},\quad\langle\tilde{Z}|\left(\left|Y\right\rangle _{1}\left|X\right\rangle _{2}\right)=\frac{1}{\sqrt{2}},\label{eq:couplcoeff}
\end{equation}
and the expressions obtained via cyclic permutation. Writing the exchange
energy as a $3\times3$ matrix with the valence band functions given
in the order $\phi_{v,\, yz}$,~$\phi_{v,\, zx}$,~$\phi_{v,\, xy}$,
we finally obtain the expression
\begin{widetext}
\begin{eqnarray}
E_{\mathrm{exch}}^{\mathrm{NA}} & = & \Delta_{Q}\frac{K^{2}}{K_{0}^{2}}\left(\begin{array}{ccc}
\hat{K}_{y}^{2}\hat{K}_{z}^{2} & \hat{K}_{z}^{2}\hat{K}_{y}\hat{K}_{x} & \hat{K}_{y}^{2}\hat{K}_{x}\hat{K}_{z}\\
\hat{K}_{z}^{2}\hat{K}_{y}\hat{K}_{x} & \hat{K}_{z}^{2}\hat{K}_{x}^{2} & \hat{K}_{x}^{2}\hat{K}_{y}\hat{K}_{z}\\
\hat{K}_{y}^{2}\hat{K}_{x}\hat{K}_{z} & \hat{K}_{x}^{2}\hat{K}_{y}\hat{K}_{z} & \hat{K}_{x}^{2}\hat{K}_{y}^{2}
\end{array}\right)\label{eq:HexchKmatrix}
\end{eqnarray}
\end{widetext}
for the nonanalytic exchange energy with $\hat{\boldsymbol{K}}=\boldsymbol{K}/K$. Contrary to dipole allowed excitons, the nonanalytic exchange
energy depends on the fourth power of the angular coordinates of $\boldsymbol{K}$. 

We can now explicitly give the coefficient
$\Delta_{Q}$ of Refs.~\cite{9_1,8,9} from microscopic calculations
and estimate its size 
using Eq.~(\ref{eq:Sexcexch}) and the values 
$\left|\boldsymbol{p}_{nm}\right|/\hbar\approx 1.3\times10^{9}\,\mathrm{m}^{-1}$
and $\left(E_{m}-E_{n}\right)\geq\Delta E=449\,\mathrm{meV}$ 
given in Ref.~\cite{1}:
\begin{align}
\Delta_{Q} = &\: \frac{6c_{\rho}^{2}e^{2}K_{0}^{2}}{\varepsilon_{0}\varepsilon_{\infty}V_{\mathrm{uc}}}\frac{\hbar^4}{m_{0}^4}|F_{vc\,1S}\left(\boldsymbol{0}\right)\sum_{n\neq v,c}g_{vc}\left(E_{n}\right)p_{nc}p_{vn}|^{2}\nonumber \\
 \approx &\: 9\,\mathrm{neV}.
\end{align}
This value is significantly smaller than the result $\Delta_{Q}=5\,\mathrm{\mu eV}$
from Ref.~\cite{9_1}. We see that also the $K$ dependent nonanalytic
exchange interaction is negligibly small in $\mathrm{Cu_{2}O}$.


As has been stated in Sec.~\ref{sec:Exchange-interaction}, it
is nevertheless interesting to investigate the possible coupling between
longitudinal and transverse exciton states. 
In general, these states are not uncoupled.
However, in the case of the ortho exciton this
restriction holds only if the $K$ vector is parallel to one of the main symmetry axes of the crystal. 
We will show that a general direction of the $K$ vector an LT coupling appears, for which reason 
all three exciton states couple to light with a polarization not
being orthogonal to the wave vector involved. 

We start with $\boldsymbol{K}$
being oriented in $\left[100\right]$ direction. In this case
the cubic symmetry is reduced to the group $C_{4\mathrm{v}}$, which
leaves $\boldsymbol{K}$ invariant. Since $K_{y}=K_{z}=0$ holds,
the nonanalytic exchange interaction~(\ref{eq:HexchKmatrix}) is
zero. Therefore, we are allowed to choose appropriate linear combinations
of the states $\phi_{v,\, yz}$,~$\phi_{v,\, zx}$,~$\phi_{v,\, xy}$
such that $\boldsymbol{\mu}_{vc\,\nu\boldsymbol{K}}\parallel\boldsymbol{K}$
and $\boldsymbol{\mu}_{vc\,\nu\boldsymbol{K}}\perp\boldsymbol{K}$
holds. To this aim, we insert the charge density 
\begin{align}
\rho_{vc,\,\nu\boldsymbol{K}}\left(\boldsymbol{r}\right) = &\: c_{\rho}e\sum_{\boldsymbol{\beta}}U_{vc\,\nu\boldsymbol{K}}\left(\boldsymbol{\beta}\right)\nonumber \\
  & \quad\times\phi_{c,\, s}\left(\boldsymbol{r}-\boldsymbol{\beta}\right)\left(c_{yz}\phi_{v,\, yz}^{*}\left(\boldsymbol{r}\right)\right.\nonumber\\
  & \quad+\left.c_{zx}\phi_{v,\, zx}^{*}\left(\boldsymbol{r}\right)+c_{xy}\phi_{v,\, xy}^{*}\left(\boldsymbol{r}\right)\right)
\end{align}
into 
\begin{equation}
\boldsymbol{\mu}_{vc\,\nu\boldsymbol{K}}=\int\mathrm{d}\boldsymbol{r}\,\boldsymbol{r}\,\rho_{vc\,\nu\boldsymbol{K}}\left(\boldsymbol{r}\right).
\end{equation}
Using Eq.~(\ref{eq:Wanfunc}) and 
considering again the coupling coefficients for the case 
$\Gamma_{4}^{-}\otimes\Gamma_{4}^{-}\rightarrow\Gamma_{5}^{+}$ [cf.
Eq.~(\ref{eq:couplcoeff})], we obtain
\begin{equation}
\boldsymbol{\mu}_{vc\,\nu\boldsymbol{K}}\sim\left(\begin{array}{c}
K_{y}c_{xy}+K_{z}c_{zx}\\
K_{z}c_{yz}+K_{x}c_{xy}\\
K_{x}c_{zx}+K_{y}c_{yz}
\end{array}\right).\label{eq:muc}
\end{equation}
Hence, the two transverse states for $\boldsymbol{K}\parallel\left[100\right]$ 
are given by $c_{xy}=1$, $c_{yz}=c_{zx}=0$ and 
$c_{zx}=1$, $c_{yz}=c_{xy}=0$. This is not unexpected
since the $K$ vector causes a symmetry breaking in $x$
direction, which affects the functions $\phi_{v,\, xy}$
and $\phi_{v,\, zx}$ in a different way than $\phi_{v,\, yz}$.

The fact that longitudinal and transverse exciton states are decoupled
for $\boldsymbol{K}\parallel\left[100\right]$ can also be understood
from group theoretical considerations: The exciton states transform
according to $\Gamma_{5}^{+}$ in $O_{\mathrm{h}}$ while the dipole
operator transforms according to the
irreducible representation $D^{1}$ of the full rotation group or
according to $\Gamma_{4}^{-}$ in $O_{\mathrm{h}}$. As the cubic
symmetry reduces to $C_{4\mathrm{v}}$, we have to consider the reduction
of the irreducible representations of the cubic group $O_{\mathrm{h}}$
by the group $C_{4\mathrm{v}}$:
\begin{subequations}
\begin{align}
\Gamma_{5}^{+} \rightarrow &\: \Gamma_{4}\oplus\Gamma_{5},\\
\displaybreak[1]
\Gamma_{4}^{-} \rightarrow &\: \Gamma_{1}\oplus\Gamma_{5}.
\end{align}
\end{subequations}
Comparing both equations, we immediately see that the two $\Gamma_{5}$ states are transverse
states and that the $\Gamma_{4}$ state is a longitudinal state. 
Since there are now exciton states transforming according to $\Gamma_{5}$
and a dipole operator which transforms according to $\Gamma_{5}$,
these exciton states can be excited by light. This describes the fact
that the $1S$ exciton becomes quadrupole allowed due to the $K$
dependent admixture of $P$ excitons. On the other hand, 
as the transverse
states and the longitudinal state
transform according to different irreducible representations, no coupling
between these states occurs.

Let us now consider the exchange interaction~(\ref{eq:HexchKmatrix})
for an arbitrary $\boldsymbol{K}$ with all vector components $K_{i}\neq0$. The eigenvalues
$\lambda_{i}$ and eigenvectors $v_{i}$ of the $3\times3$ matrix
in Eq.~(\ref{eq:HexchKmatrix}) read
\begin{align}
\lambda_{1}=0, \quad &\: v_{1}=\frac{\left(-b,\, a,\,0\right)^{\mathrm{T}}}{\sqrt{a^{2}+b^{2}}},\\
\displaybreak[1]
\lambda_{2}=0, \quad &\: v_{2}=\frac{\left(-c,\,0,\, a\right)^{\mathrm{T}}}{\sqrt{a^{2}+c^{2}}},\\
\displaybreak[1]
\lambda_{3}=a^{2}+b^{2}+c^{2}, \quad &\: v_{3}=\frac{\left(a,\, b,\, c\right)^{\mathrm{T}}}{\sqrt{a^{2}+b^{2}+c^{2}}},
\end{align}
with the abbreviations $a=\hat{K}_{y}\hat{K}_{z}$, $b=\hat{K}_{z}\hat{K}_{x}$
and $c=\hat{K}_{x}\hat{K}_{y}$. Even though there is only one state
with an eigenvalue $\lambda\neq0$, we have to prove that this state
is connected with a longitudinal polarization. Inserting $v_{3}$
into Eq.~(\ref{eq:muc}) yields
\begin{equation}
\boldsymbol{\mu}_{vc\,\nu\boldsymbol{K}}\sim\boldsymbol{K}-K\left(\hat{K}_{x}^{3},\,\hat{K}_{x}^{3},\,\hat{K}_{x}^{3}\right)^{\mathrm{T}}
\end{equation}
Due to the second term, the dipole moment is not parallel to $\boldsymbol{K}$.
Therefore, we have shown that longitudinal and transverse exciton
states are coupled by the nonanalytic exchange interaction~(\ref{eq:HexchKmatrix})
if $\boldsymbol{K}$ is not oriented in a direction of high symmetry.
Furthermore, we see that two eigenstates of the matrix in Eq.~(\ref{eq:HexchKmatrix})
are degenerate. If the nonanalytic exchange interaction were the only
reason for the $K$ dependent splitting of the $1S$ exciton, only
two states would be observable in experiments for any direction of
$\boldsymbol{K}$.

\section{Summary and outlook~\label{sec:Summary-and-outlook}}

Using $\boldsymbol{k}\cdot\boldsymbol{p}$ perturbation theory,
we could derive 
$K$ dependent higher order terms of the analytic
and nonanalytic exchange interaction
of Wannier excitons. We have discussed the specific
properties of $\mathrm{Cu_{2}O}$ and in particular the effects
of the valence band structure. Investigating the $K$ dependent exchange interaction
of the $1S$ excitons in this semiconductor, we could show that the
$K$ dependent terms of the analytic and the nonanalytic exchange
interaction are negligibly small compared to the 
effects of the nonisotropic dispersion.
A closer examination of the $K$ dependent nonanalytic exchange interaction exhibited 
a coupling between longitudinal and transverse exciton states
if $\boldsymbol{K}$ is not oriented in a direction of high symmetry.


%

\end{document}